\documentclass{JINST}

\usepackage{amssymb,amsmath}
\usepackage{amsthm}
\usepackage{graphicx}

\title{Results for the response function determination of the Compact Neutron Spectrometer}

\author{F. Gagnon-Moisan$^a$\thanks{Corresponding
author.}~, M. Reginatto$^a$ and A. Zimbal$^a$\\
\llap{$^a$}Physikalisch-Technische Bundesanstalt,\\
   Bundesallee 100, D-38116, Braunschweig, Germany\\

  E-mail: \email{francis.gagnon-moisan@ptb.de}}

\abstract{The Compact Neutron Spectrometer (CNS) is a Joint European Torus (JET) Enhancement Project, designed for fusion diagnostics in different plasma scenarios. The CNS is based on a liquid scintillator (BC501A) which allows good discrimination between neutron and gamma radiation. Neutron spectrometry with a BC501A spectrometer requires the use of a reliable, fully characterized detector. The determination of the response matrix was carried out at the Ion Accelerator Facility (PIAF) of the Physikalisch-Technische Bundesanstalt (PTB). This facility provides several monoenergetic beams (2.5, 8, 10, 12 and 14 MeV) and a \textit{white field }($E_{max} \sim 17$ MeV), which allows for a full characterization of the spectrometer in the region of interest (from $\sim 1.5$ MeV to $\sim 17$ MeV). The energy of the incoming neutrons was determined by the time of flight method (TOF), with time resolution in the order of 1 ns. To check the response matrix, the measured pulse height spectra were unfolded with the code MAXED and the resulting energy distributions were compared with those obtained from TOF. The CNS project required modification of the PTB BC501A spectrometer design, to replace an analog data acquisition system (NIM modules) with a digital system developed by the \textit{Ente per le Nuove tecnologie, l'Energia e l'Ambiente} (ENEA). Results for the new digital system were evaluated using new software developed specifically for this project.
}

\begin{document}
%% main text
\section{Introduction}\label{intro}

Fusion facilities like the Joint European Torus (JET) require sophisticated diagnostic tools to effectively monitor the plasma inside the reactor torus. Numerous studies have been carried out using different types of apparatus, from simple optical devices to very sophisticated detector systems \cite{hutch2005}. Still, to properly evaluate the reaction rate one needs to detect the products of the reaction; i.e., neutrons. Because neutrons interact little with matter, they are good candidates for properly characterizing the plasma. There are two neutron producing reactions which are of interest, the d+D and d+T reactions, which produce 2.45 and 14.1 MeV neutrons respectively.

The Compact Neutron Spectrometer (CNS) project is a JET Enhancement Project for neutron fusion diagnostics in different plasma scenarios. It is based on a liquid scintillator (BC501A) which allows for good discrimination of neutron (n) and gamma ($\gamma$) radiation. The spectrometer is able to detect neutrons over an energy range that covers both of the neutron energies mentioned above. To do neutron spectrometry with a BC501A spectrometer, it is necessary to have a reliable, fully characterized detector: it is crucial to have a good response matrix and, when operating under challenging circumstances such as the ones encountered at JET, the means to correct for the effect of any environmental conditions that cannot be effectively eliminated by shielding the detector. The \textit{Physikalisch-Technische Bundesanstalt} (PTB) has many years of experience in the development, construction and use of liquid scintillation neutron detectors as neutron spectrometers \cite{Klein2002}. The BC501A neutron spectrometer developed at the PTB has been shown to meet the requirements for fusion diagnostics \cite{Zimbal2004,Reginatto2008}. A system similar to the one described in this paper has been built by PTB and installed at ASDEX upgrade \cite{Tardini2012}. The procedures for the determination of the neutron response matrix described here will be applied to the ASDEX upgrade neutron spectrometer as well.

The experimental characterization of the CNS has been carried out at the PTB accelerator facility \cite{Nolte2004}. This experimental facility provides unique neutron radiation fields in the energy range from a few keV to 19 MeV. For the CNS project, a new method of data analysis had to be developed on account of the new digital acquisition system (DPSD) developed by the \textit{Ente per le Nuove tecnologie, l'Energia e l'Ambiente }(ENEA) which replaced the analog data acquisition system (NIM modules) that was previously in use \cite{Moisan2012}. The software necessary for the data analysis was developed at PTB.

\section{CNS Project}\label{CNSP}

The objectives of the CNS project can be summarized as follows:
\begin{itemize}
\item Construction of the CNS at PTB and verification of the performance requirements.
\item Development of a computer code to carry out the analysis. The code was written in C++ and it makes use of the ROOT framework \cite{root}.
\item Characterization of a BC501A fast-neutron detector with a digital acquisition system developed by the ENEA group.
\item Determination of two response matrices (RM), an experimentally determined response matrix and a response matrix based on simulations adjusted to fit the data collected during the measurement campaign.
\item Evaluation of the resolution that is achievable after unfolding the measurements with the MAXED code, and comparison with Time-Of-Flight (TOF) measurements.
\end{itemize}
Using unfolding techniques, the CNS should be able to achieve a resolution ($\Delta E/E$) of 6\% for the 2.5 MeV peak and 4\% for the 14 MeV neutron peak. %\cite{Zimbal2004}.
It should also be able to handle the high count rates that are expected at JET \cite{Marocco2009}.

\section{Compact neutron spectrometer}

\begin{figure}[!ht]
\centering
\includegraphics[width=4.0in]{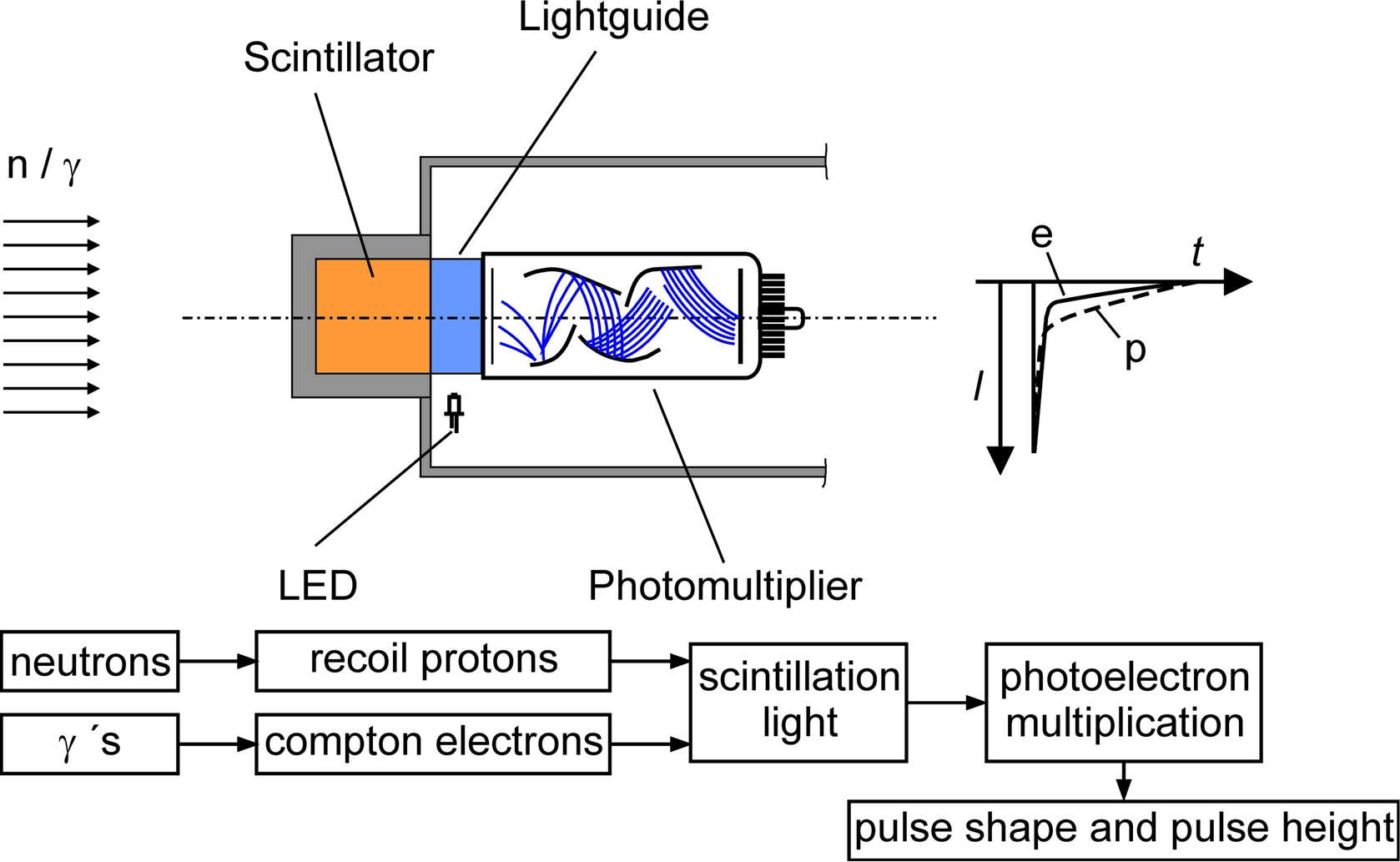}
\caption{A diagram of the compact neutron spectrometer, illustrating the operating principle.}
\label{fig:CNS}
\end{figure}

PTB has developed the CNS based on the requirements of the JET project. It consist of a BC501A standard cell of small size from Bicron ($\emptyset$ = 2.5 cm, $h$ = 2.5 cm) coupled to a Phillips XP2020 photomultiplier tube (PMT) via a light guide. The cell is assembled in a sealed aluminium housing. A pulsed LED is integrated into the design and used for gain correction of the PMT. A diagram of the detector is presented in Fig. \ref{fig:CNS}.

The CNS makes use of a digital acquisition system made by ENEA to satisfy JET needs \cite{Marocco2009, Esposito2004}. This board allows full digitalization of the PMT signal, with a sampling rate of 200 Msamples and 14 bit/ch resolution. The full digitalized pulses are recorded on a hard drive, which allows the analysis to be carried out after the experiment. The board also provides a TOF mode, using the PMT signal as start trigger. Neutron and $\gamma$ induced events can be separated using pulse-shape discrimination (PSD) techniques  \cite{Zimbal2004, Marocco2009, Roush1964, Knoll2010}.

\subsection{Pulse height resolution and light output function}
\label{sec:PHR}

The scintillator used for the CNS, BC501A (also known as NE213), is an organic compound made from xylene ($C_{8}H_{10}$) and naphthalene ($C_{10}H_{8}$). The light output due to neutron interactions in the material has been studied at PTB over the past 30 years \cite{Brooks2002}. The main light component comes from the slowing down of elastic scattered protons, but light can also be produced by the recoil of $\alpha$-particles, electrons, or carbon nuclei. The light output is normalized to the light produced by a 1 MeV electron (MeV$_{\mathrm{ee}}$).

The Monte Carlo code NRESP \cite{NRESP} has been used successfully at PTB to simulate pulse height spectra (PHS) for a given BC501A cell geometry and neutron energy. The simulated PHS calculated by NRESP have a sharp edge which in actual measurements is smoothed out. The simulated PHS therefore have to be adjusted to fit the measured PHS, and this is done using Gaussian broadening (further adjustments that are also required for a good fit are discussed in section \ref{adjsimresp}). The amount of Gaussian broadening defines the intrinsic pulse height resolution of the spectrometer. The pulse height function $L(E)$ and the pulse height resolution function $\mathrm{d}L/L$ are characteristic of each individual detector. Eq.(\ref{eq:Eresol}) gives a formula for $\mathrm{d}L/L$ in terms of fit parameters A, B and C \cite{Klein2002},
\begin{equation}
\left( \frac{\mathrm{d}L}{L} \right)  = \sqrt{A^{2}+\frac{B^{2}}{L}+\left( \frac{C}{L}\right) ^{2}}.
\label{eq:Eresol}
\end{equation}
The parameters can be associated with contributions that originate from different processes: the light transmission from the scintillator to the photocathode ($A$), the statistical nature of the light production ($B$), and electronic noise ($C$).

\section{PTB facility}

\begin{figure}[!ht]
\centering
\includegraphics[width=3.5in]{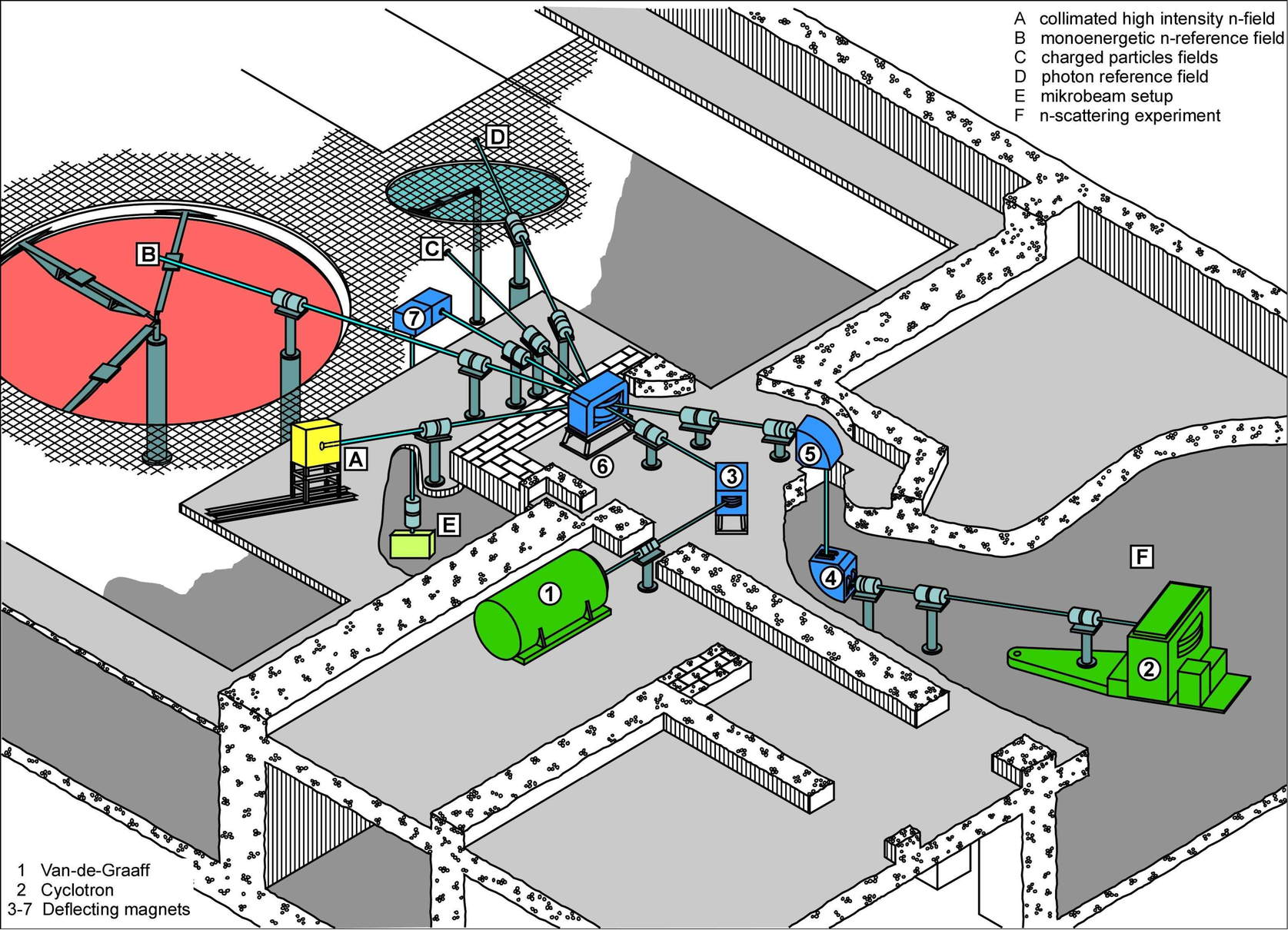}
\caption{The PTB experimental area, with the van de Graaff accelerator and the cyclotron }
\label{fig:PTB}
\end{figure}

The measurements needed for the characterization of the CNS were carried out at the PTB Ion Accelerator Facility (PIAF) \cite{Nolte2004} (see Fig. \ref{fig:PTB}). Several monoenergetic neutron beams (8, 10, 12 and 14 MeV) and a broad neutron field ($E_{\mathrm{max}}\;\sim\;$17 MeV, also called \textit{white field}), produced using a cyclotron, are available. This allows a full characterization of the spectrometer in the region of interest (from $\sim$ 1.5 MeV to $\sim$ 17 MeV).

The PTB neutron reference radiation fields used for these measurements are produced using D(d,n) and $^{9}$Be(p,n) reactions. The first reaction is produced by means of a deuterium gas target. The beam has a small energy loss in the target (typically less than $\sim\:100$ keV) and is fully stopped in the backing yielding quasi-monoenergetic neutron beams. The second reaction is produced by means of a thick 3 mm beryllium target, which is thick enough to fully stop the incoming protons yielding a neutron beam with a broad energy spectrum (white field).

The main experimental campaign was carried out in two steps; the first involved the measurement of four monoenergetic beams (8, 10, 12 and 14 MeV), the second a measurement in the white field for approximately 72 hours. The selector frequency of the cyclotron was $\sim$ 900 kHz, with a current of $I=900$ nA on the deuterium gas target. The detector was placed at a distance of 12.29 m from the target. The uncertainty in the distance determination is assumed to be $\pm$ 1 cm.

In addition to the main campaign, a second campaign was carried out with neutron reference fields generated with the Van der Graaff accelerator. These neutron fields are produced with a tritiated titanium (Ti(T)) target, using the reaction $^{3}$H(d,n)$^{4}$He for the 14 MeV field and the reaction $^{2}$H(d,n)$^{3}$He for the 2.5 MeV field. With measurements in these fields, it is possible to obtain an independent test of the neutron response matrix and to verify the neutron energy resolution after unfolding as the neutron energies of these fields (for some particular reaction conditions) are not determined by the properties of the targets or the ion beams \cite{Zimbal2006}.

\section{Method of analysis}

\subsection{Time of flight method}\label{tof}

In the time of flight method, the kinetic energy of a particle is determined using measurements of the travel time over a fixed distance determined by two reference points \cite{Knoll2010, Schmidt1998}.

\begin{figure}[ht]
\centering
\includegraphics[width=5in]{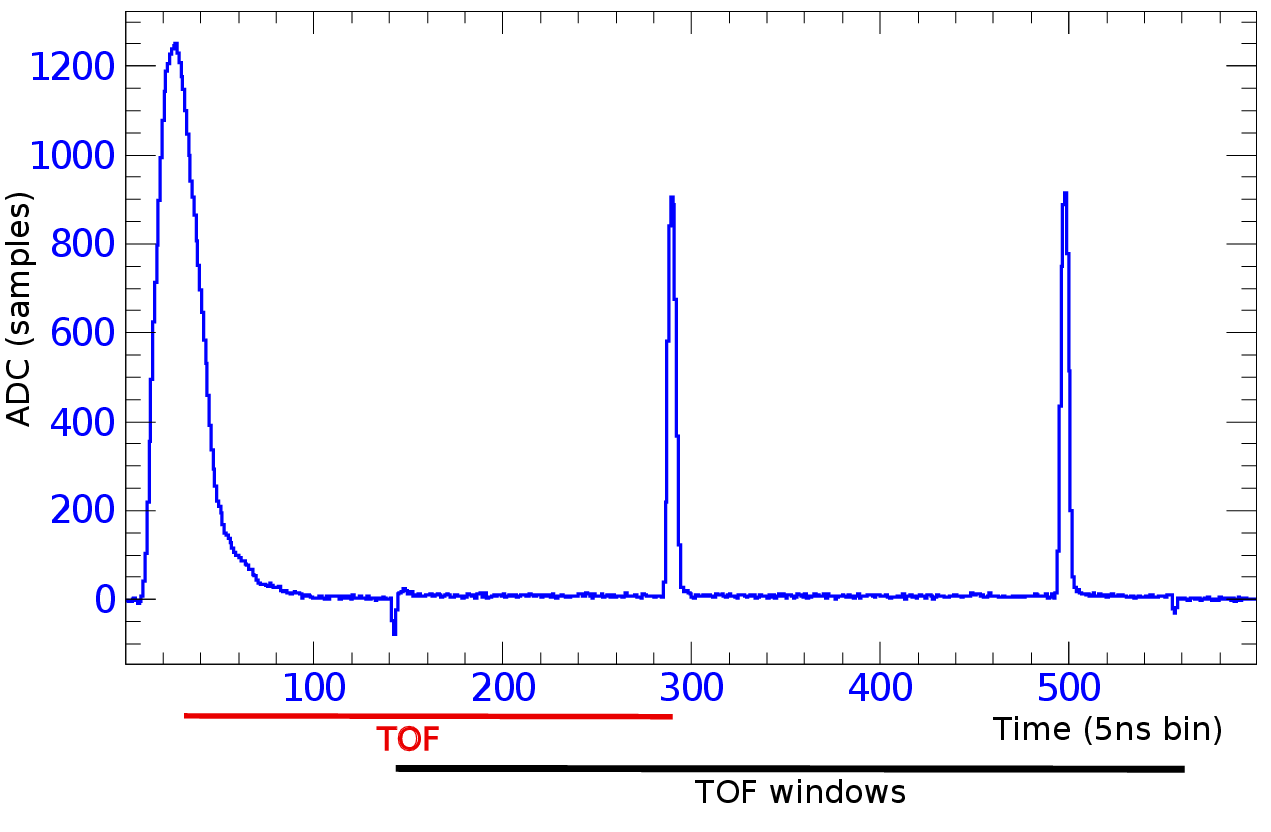}
\caption{A pulse from the PMT together with cyclotron signals recorded by the ENEA DPSD system. The pulse from the PMT is on the left. The TOF window starts after 120 samples (each sample is 5 ns). The signal polarity is inverted.}
\label{fig:pulse}
\end{figure}

The cyclotron provides a signal for each particle bunch that is emitted. When the acquisition board records a pulse from the PMT, the acquisition switches from the PMT to the signal from the cyclotron after a pre-defined amount of samples. A small negative pulse, caused by a difference in the baseline values of the two input channels of the fast filter/switch unit, indicates the switch between signals. The cyclotron signal is recorded and it shows up as a peak for each particle bunch emitted from the accelerator. Fig. \ref{fig:pulse} shows an example of this procedure.

\begin{figure}[ht]
\centering
\includegraphics[width=5in]{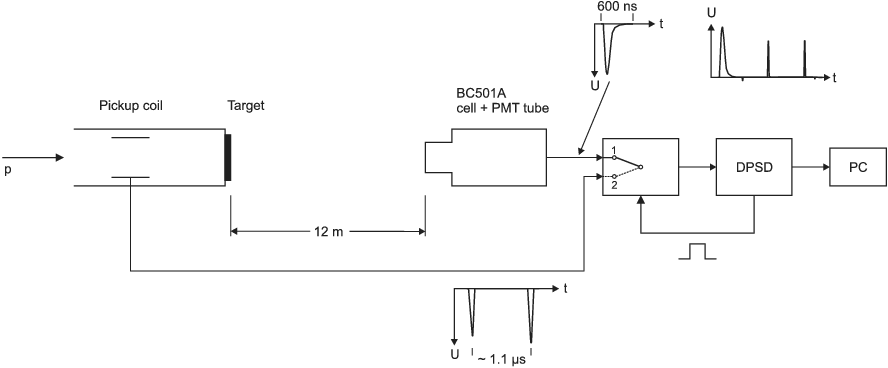}
\caption{Principle of data acquisition for TOF determination. If an event is identified as a signal over threshold by the DPSD system, an output from the DPSD board is used to switch the input of a fast analog switch device (PTB design). This switch is delayed 500 - 600 ns to have sufficient time to sample the complete event (neutron, gamma, or LED pulse). After the event, the reference signal from the pickup coil is sampled for 2 - 3 $\mu$s. Finally, the switch goes back to input 1 and the system is ready for the next event.}
\label{fig:DPSD-Filter}
\end{figure}

The schematic layout of this procedure, which follows \cite{Bardelli2007}, is shown in Fig. \ref{fig:DPSD-Filter}. Note that the negative pulses from the PMT and reference signals are inverted by the DPSD acquisition system.

The signals with 5 ns sampling time were interpolated to 1 ns steps and a 40\% digital constant fraction method was used to determine the peak position for TOF evaluation, both for the pulse of the PMT and for the reference pulses from the cyclotron. The time difference between the PTM signal pulse and the first cyclotron reference pulse is the TOF value of the event. The time scale is based of the sampling frequency of the ADC (5 ns/sample). The time distance between 2 cyclotron reference pulses determined in the same way is the inverse of the repetition frequency of the cyclotron, which is determined by comparison with a standard frequency traceable to the PTB atomic clock. Using this procedure, the sampling frequency of the ADC was confirmed to be 5 ns with a relative uncertainty of $1.5\times 10^{-4}$.

\begin{figure}[ht]
\centering
\includegraphics[width=4.5in]{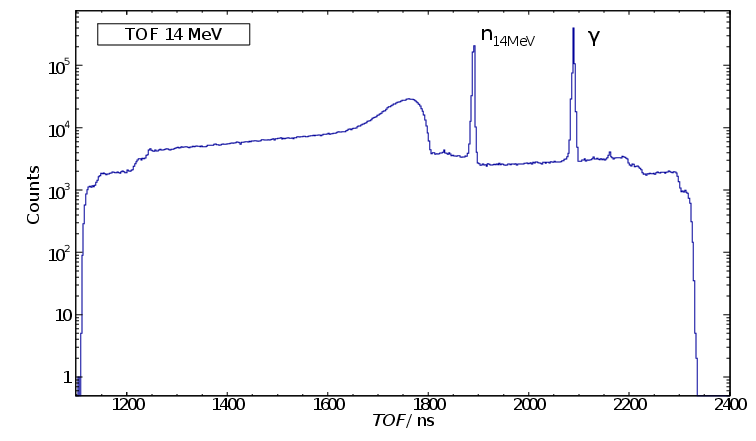}
\caption{Time-of-flight measurement of a 14 MeV monoenergetic neutron field produced at the PTB cyclotron (distance of 12.29 m from the target), showing (from right to left) the gamma peak, the 14 MeV neutron peak and the broad distribution of breakup neutron.}
\label{fig:tof14}
\end{figure}

Fig. \ref{fig:tof14} shows the time distribution of events recorded for measurements in a neutron field of 14 MeV. For the fastest particles (i.e., photons), the TOF value is higher because the PMT pulse is used as the start signal. The calibration of the zero point of the TOF scale for neutrons is done using the TOF of photons, which travel at a fixed velocity. The time resolution achieved with this methods is shown by the time resolution achieved for the gamma signals (the FWHM of the gamma TOF peak is 2.8 ns). This proves that the sample frequency of 200 MHz that was used for the experiment is adequate for our application. From the uncertainty in the time determination and the assumed uncertainty in the distance of 1 cm, an uncertainty in the neutron energy of 200 keV at 14 MeV can be estimated. 

\subsection{n-$\gamma$ discrimination}

\begin{figure}[ht]
\centering
\includegraphics[width=4in]{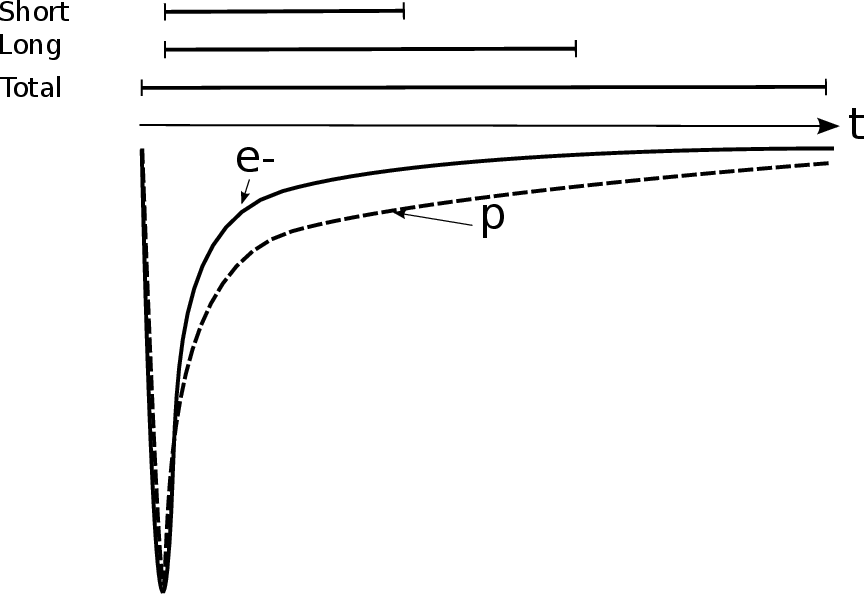}
\caption{Pulse shapes for electrons and protons. Different integration gates are also shown.}
\label{fig:ppmt}
\end{figure}

The shape of the light emission from BC501A is different for signals due to neutron and gamma radiation, as shown on Fig. \ref{fig:ppmt}. This permits the use of pulse-shape discrimination (PSD) techniques to distinguish between these two types of radiation. Events are discriminated using this technique and only neutron events are used for constructing the response matrix.

\begin{figure}[ht]
\centering
\includegraphics[width=5in]{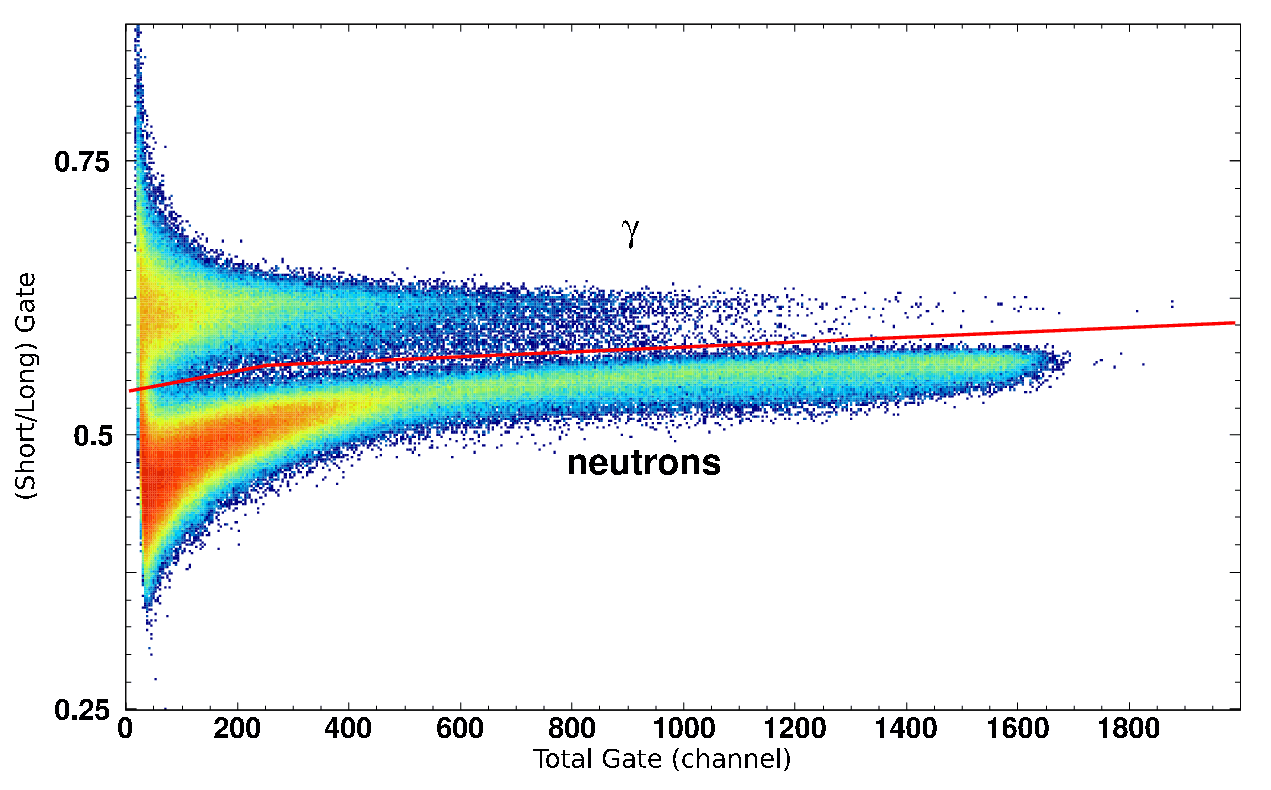}
\caption{PSD identification plot. The line drawn in red separates $\gamma$ events (top) from neutron events (bottom). For this example, data from measurements made in the 12 MeV monoenergetic neutron field were used.}
\label{fig:psd}
\end{figure}

The PSD method consists of defining short, long, and total integration gates (see Fig. \ref{fig:ppmt}), taking the ratio of the short integration gate to the long integration gate, and plotting this value against the total integration gate, as shown in Fig. \ref{fig:psd}. For the CNS, a value of 3/14 was selected for the short over long gate ratio. The PSD procedure has been implemented by the ENEA group with the digital board analysis software (DPSD) \cite{Zimbal2004, Marocco2009,Esposito2004}.

\subsection{Satellite correction}

\begin{figure}[ht]
\centering
\includegraphics[width=4.5in]{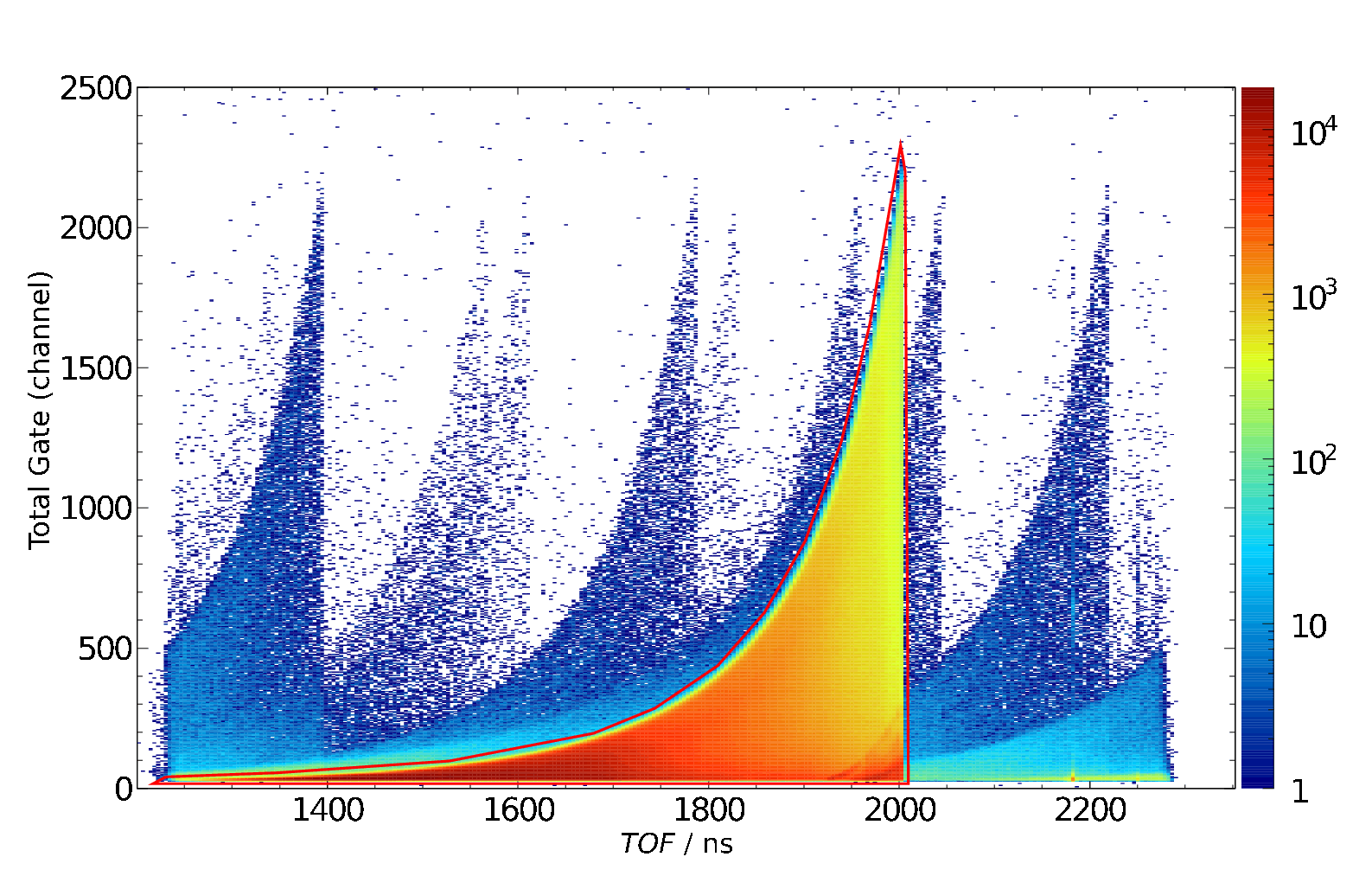}
\caption{Pulse height versus TOF distribution for measurements in the white field. Only events identified as neutron are displayed. The red contour indicates the selection of the events with correct time evaluation. Events from other pulses (satellites) that overlap with this region are removed using a stochastic procedure.}
\label{fig:sat}
\end{figure}

The reaction $^{9}$Be(p,n) is used to obtain the white field. When a pulsed proton beam is produced, some of the proton bunches are not completely deflected by the internal beam pulse selector. This creates \textit{ghost} images of the main data structure, commonly known as satellite peaks, in the pulse height versus TOF distribution. An illustration of this situation is given in Fig. \ref{fig:sat}. These events (in the order of 7\% of the total number of events) need to be removed from the analysis because their TOF cannot be evaluated properly. This is achieved using a stochastic procedure in which a fraction of events are removed at random according to the probability that an event may belong to a satellite peak.

\subsection{Adjustment of simulated response functions and pulse height resolution}\label{adjsimresp}

We use an iterative procedure to adjust the simulated response functions (i.e., the PHS that result from the simulation) and determine the light output function of the CNS. Each iteration consists of two steps. The first step is to calculate simulated response functions for neutrons with the code NRESP using an estimate of the light output function (this estimate comes from known light output functions for similar detectors). The second step is to adjust the simulated response functions to fit the measured data (using the procedure that is described in detail below), which results in a corrected, improved light output function that can be used for a new calculation if needed. This iterative procedure converges quickly (in the case of the CNS, only two iterations were needed) and the end result is a set of corrected simulated response functions that fit the measured data well.

We now describe the procedure used to adjust the simulated response functions. For each neutron energy, the simulated PHS needs to be scaled in the x-axis (to account for differences in the gain assumed in the simulation and the gain used for the measurement) as well as in the y-axis (to account for differences between the neutron fluence assumed for the simulation and the neutron fluence of the measurement). It is also necessary to adjust the simulated PHS to allow for a difference due to a zero offset in the measurement (the zero offset is the same for all neutron energies). In addition to the calculated and measured neutron PHS, gamma PHS (e.g., from $^{207}$Bi) are needed to determine the common PHS energy scale in equivalent electron energies. The PHS for scintillation detectors from photons can be calculated with the PHRESP code \cite{Novotny1997}. $^{207}$Bi is a gamma source that emits three different, well separated gamma lines. By comparison to a gamma PHS measured under the same experimental conditions that apply to the neutron data and a calculated gamma PHS, the electron equivalent energy scale and the common zero offset can be determined. This is achieved by including the measured and calculated gamma PHS into the analysis to determine the zero offset for all the PHS. A change in the neutron light output function shows up as a residual difference between the gamma and neutron x-axis stretching factors. The simulated PHS needs to be Gaussian broadened, where the Gaussian broadening is expressed by Eq. (\ref{dLoverL}).

In previous works, the fitting procedure used to adjust the simulated PHS was sequential; i.e., one neutron energy at the time, starting with the highest neutron energy available. In the new method that is applied here, all the parameters are determined simultaneously, using a maximization procedure based on the relative entropy calculated between measured PHS and corrected simulated PHS; i.e., the set of parameters that maximizes this relative entropy is the optimal set of parameters which gives the best fit. We used the relative entropy $S$ in the form due Skilling \cite{Linden1995, Sivia2006},
\begin{equation}
\mathrm{S}(\phi^{\mathrm{s}},\phi^{\mathrm{e}}) = \sum_{\mathrm{i_{min}}}^{\mathrm{i_{max}}}\phi^{\mathrm{s}}_{\mathrm{i}}\ln\left( \frac{\phi^{\mathrm{s}}_{\mathrm{i}}}{\phi^{\mathrm{e}}_{\mathrm{i}}}\right) -(\phi^{\mathrm{s}}_{\mathrm{i}}-\phi^{\mathrm{e}}_{\mathrm{i}}),
\label{eq:Entropy}
\end{equation}
where $\mathrm{i_{min}}$ and $\mathrm{i_{max}}$ are the minimum and maximum bins of the measured and corrected simulated PHS, $\phi^{\mathrm{e}}_{\mathrm{i}}$ and $\phi^{\mathrm{s}}_{\mathrm{i}}$. The values of $\mathrm{i_{min}}$ and $\mathrm{i_{max}}$ are selected for each energy and correspond to the region of the edge on the pulse height spectrum. The maximization is carried out with respect to the parameters that are used to adjust the simulated spectrum. To carry out the maximization, we used the Broyden-Fletcher-Goldfarb-Shanno (BFGS) algorithm \cite{fletcher} that is implemented in the GNU Scientific Library (GSL) \cite{GSL}. More details of this particular application of relative entropy maximization can be found in \cite{Moisan2012}.

In previous works, the adjustment of the simulated response functions was only carried out for those energies for which there were measurements of monoenergetic beams available. In the new method used here, we use the TOF information of events from the measurement in the white field to select only those events that correspond to a given energy (more precisely, within a small energy interval about the given energy). This is possible because all the required information is stored by the digital acquisition. A similar method has been used previously but with an analog acquisition system \cite{Zimbal2006}.

For the characterization of the CNS, 30 neutron energies between 1.5 MeV and 17 MeV were selected. The energy width of each selection was chosen taking into consideration the energy binning that was used for the NRESP calculation and the counting statistics available. A width that corresponded to two of the energy bins used for the NRESP calculation was selected in order to have enough statistics in each of the measured PHS. The energy binning used was non-linear, it varied from $\Delta E_{2.5} = 23$ keV at 2.5 MeV to $\Delta E_{14} = 66$ keV at 14 MeV.

\begin{figure}[ht]
\centering
\includegraphics[width=4.5in]{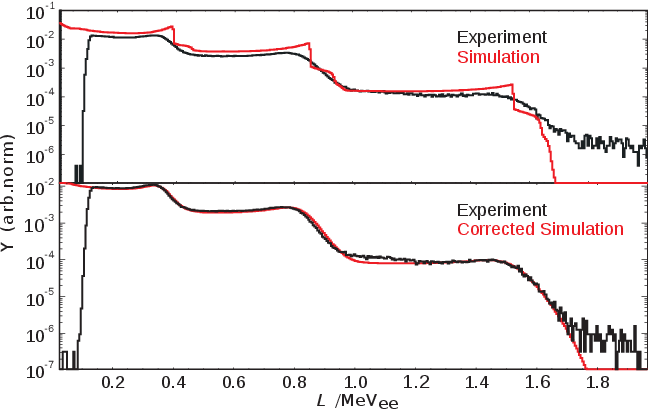}
\caption{The upper figure shows the measured PHS from a $^{207}Bi$ $\gamma$-source (black) and the PHS simulated using PHRESP (red). The lower figure shows the corrected simulated PHS (red) ploted with the same measured PHS (black) shown in the upper figure.}
\label{fig:gb}
\end{figure}

The software written to carry out the fitting procedure provides a real-time display of the maximization status and, if needed, the user can easily restart the maximization in case of non-convergence. An example of a fit for a gamma PHS from a $^{207}$Bi $\gamma$-calibration source is shown in Fig. \ref{fig:gb}.

\subsection{Experimentally determined response functions}

\begin{figure}[ht]
\centering
\includegraphics[width=4.5in]{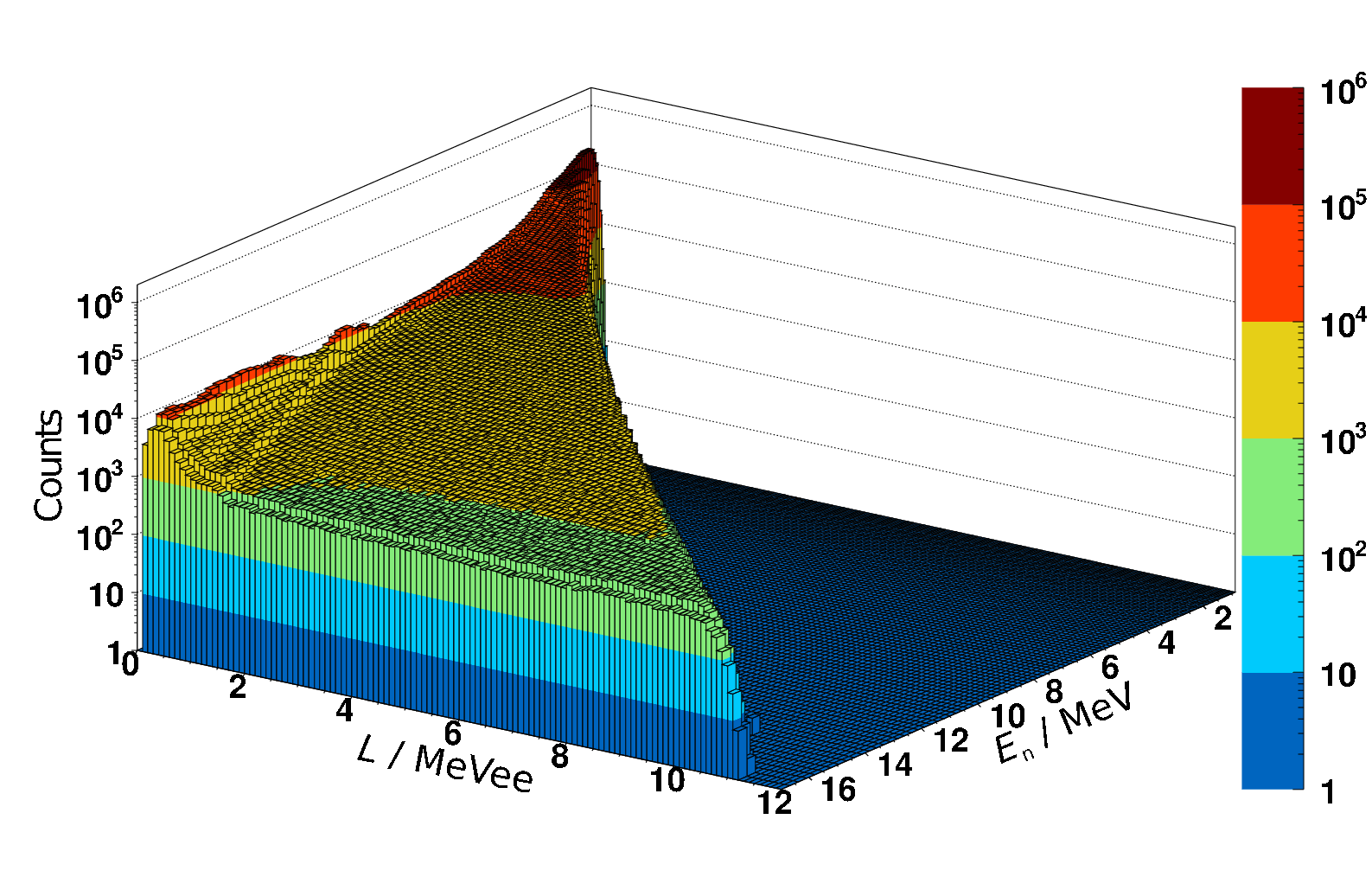}
\caption{Data used for the determination of the experimental response matrix (to reduce the number of matrix elements plotted in the figure, a coarse binning has been used where 8 neutron energy bins and 20 PHS bins have been combined). A total of 28$\times10^{6}$ neutron events were evaluated.}
\label{fig:exprm}
\end{figure}

The experimental response matrix is put together from measurements made in the white field, by selecting data from particular neutron energy ranges. Fig. \ref{fig:exprm} shows all neutron events which were analyzed (28$\times10^{6}$ events). The energy ranges chosen correspond to the neutron energy bins used for the NRESP simulations \cite{Zimbal2006}. To correct for low measurement statistics, an additional step is carried out and smoothing is introduced.

\subsection{Unfolding of measurements of monoenergetic beams and energy resolution}

The PHS from measurements of monoenergetic neutron beams were unfolded using the code MAXED together with the experimental and the corrected simulated response matrices. This resulted in neutron energy spectra for the monoenergetic neutron beams. To determine the neutron energy resolution, we compared the FWHM of simulations of the neutron energy beams with the FWHM of the unfolded spectra. We also compared to the FWHM determined using TOF. The results are presented in the next section.

\section{Results}

\subsection{Pulse height resolution}

\begin{figure}[ht]
\centering
\includegraphics[width=4.5in]{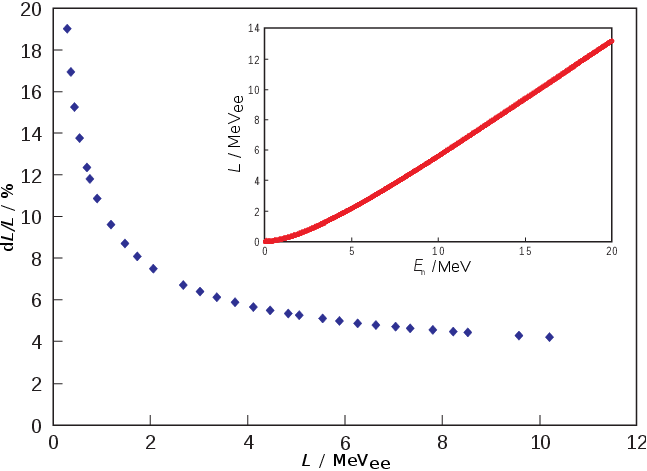}
\caption{Pulse height resolution  function $\mathrm{d}L/L$. The inset shows the relation between incoming neutron energy $E_{n}$ and light output.}
\label{fig:ResPHS}
\end{figure}

The procedure used to adjust the simulated spectra gives the following values for the Gaussian broadening parameters: $A=2.82$, $B=10.0$ and $C=0.6$. A plot of the light output resolution (see eq. \ref{eq:Eresol}) is shown in Fig. \ref{fig:ResPHS}.

The set of parameters that regulates the scaling of the simulated PHS in the x-axis for different neutron energies (see the discussion in subsection \ref{adjsimresp}) is also of interest. When the light output function used for the NRESP simulation is adequate, this set of scaling parameters should differ from 1 by less than a percent. This is indeed the case after the last iteration of our procedure is carried out: At $E_{\mathrm{n}}=2.5$ MeV and $E_{\mathrm{n}}=14$ MeV, the difference from 1 is -0.90\% and +0.35\% respectively.

\subsection{Measurements in the white field}

\begin{figure}[!t]
\centering
\includegraphics[width=5.5in]{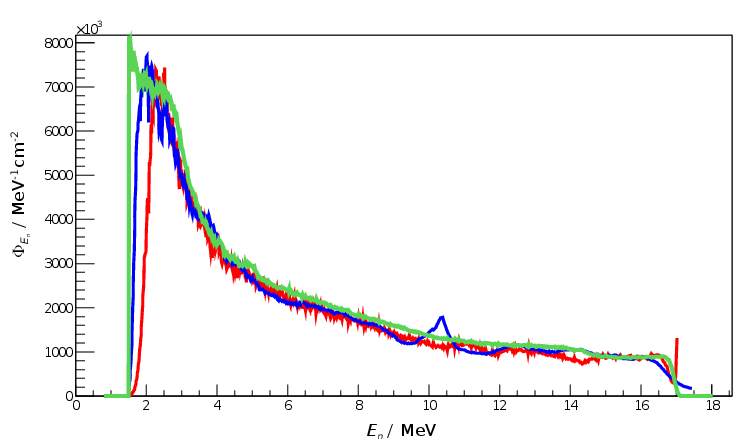}
\caption{Measured spectral neutron fluence of the broad neutron field unfolded using the experimental (in red) and corrected simulatred (in blue) response matrices. These unfoldings are compared to the experimental evaluation of TOF (in green).}
\label{fig:UnfWF}
\end{figure}

The white neutron field provides a good test for the CNS, since it is a continuous spectrum that covers all neutron energies to up to $\sim$ 17 MeV. Fig. \ref{fig:UnfWF} shows unfoldings carried out using both experimental and corrected simulated responses, together with the result obtained using the TOF evaluation.

\subsection{Mono-energetic beams}

Table \ref{tab:res} shows a comparison of the three different methods of estimating the mean energy and the FWHM of the monoenergetic neutron beams that were measured during the experimental campaign. For the first set of measurements, which were carried out using the cyclotron, TOF information is available. The last 2 measurements were carried out with the Van de Graaff accelerator, and no TOF information is available. For the last two lines, we provide estimates of the mean energy and the FWHM derived from simulations done with the code TARGET \cite{TARGET} (these values are indicated with an asterisk).

\begin{figure}[ht]
\centering
\includegraphics[width=4.5in]{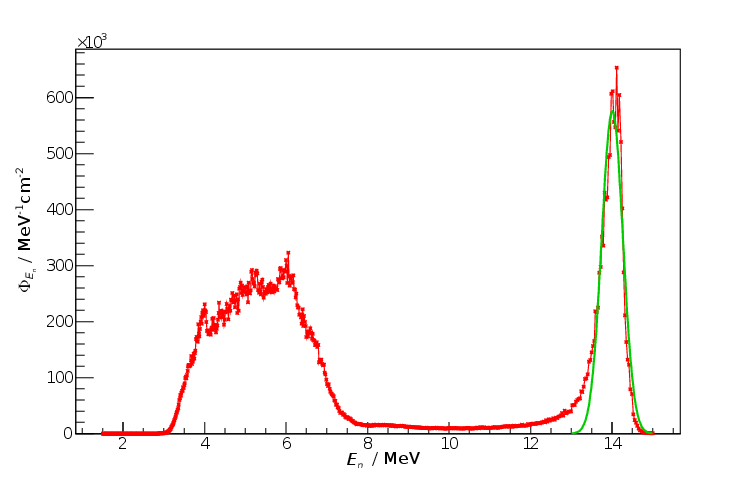}
\caption{Unfolding of the 14 MeV monoenergetic field (red), showing both the main energy peak at 14 MeV and the break-up process at lower energies. The green curve is the Gaussian fit used to extract the mean energy and FWHM of the peak. }
\label{fig:En14MeV}
\end{figure}

\begin{figure}[ht]
\centering
\includegraphics[width=4.5in]{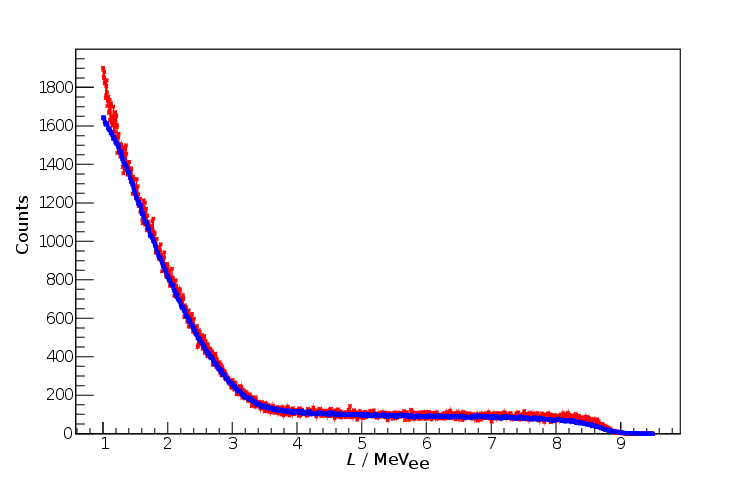}
\caption{Measured PHS (red) compared to the predicted PHS (blue) that results from folding the neutron energy spectrum derived with the unfolding with the response functions.}
\label{fig:ph14MeV}
\end{figure}

In general, the unfoldings were done using a range that excluded the lower energy channels of the PHS but large enough to result in an unfolding that showed the complete peak. The one exception is the unfolding that corresponds to the values in the first line (in italics), which extends to much lower PHS channels so that neutrons that result from the D break-up in the d(D,n) reaction for neutron energies > 8 MeV also appear in the unfolded spectrum (see Figs. \ref{fig:En14MeV} and \ref{fig:ph14MeV}).

\begin{table}
\begin{tabular}{ccc|ccc|ccc}
\hline
 \textbf{$E_{\mathrm{n}}^{exp}$} & \textbf{$FWHM^{exp}$} & \textbf{$Res^{exp}$} & \textbf{$E_{\mathrm{n}}^{sim}$} & \textbf{$FWHM^{sim}$} & \textbf{$Res^{sim}$} & \textbf{$E_{\mathrm{n}}^{TOF}$} & \textbf{$FWHM^{TOF}$} & \textbf{$Res^{TOF}$} \\
\textbf{(MeV)} & \textbf{(MeV)} & \textbf{(\%)} & \textbf{(MeV)} & \textbf{(MeV)} & \textbf{(\%)} & \textbf{(MeV)} & \textbf{(MeV)} & \textbf{(\%)} \\
\hline
\textit{\textbf{14.01} }& \textit{0.62}& \textit{4.4} & \textit{\textbf{14.10}} & \textit{0.50} &\textit{3.5} & \textit{\textbf{14.08}} & \textit{0.31} & \textit{2.2} \\
\textbf{14.03} & 0.37 & 2.6 & \textbf{14.11} & 0.22 & 1.6 & \textbf{14.08} & 0.31 & 2.2 \\
\textbf{11.99} & 0.32 & 2.6 & \textbf{12.08} & 0.20 & 1.7 & \textbf{12.04} & 0.20 & 1.6 \\
\textbf{10.03} & 0.28 & 2.8 & \textbf{10.13} & 0.19 & 1.9 & \textbf{9.99}  & 0.19 & 1.9 \\
\textbf{8.09 } & 0.22 & 2.7 & \textbf{8.16}  & 0.20 & 2.5 & \textbf{8.08}  & 0.24 & 2.9 \\
\hline
\textbf{2.36 } & 0.10 & 4.4 & \textbf{2.38}  & 0.14 & 6.0 & \textbf{*2.35}  & *0.09  & *3.6 \\
\textbf{13.94}  & 0.29 & 2.1 & \textbf{14.04} & 0.17 & 1.2 & \textbf{*13.97} & *0.20 & *1.4 \\
\hline
\end{tabular}
\caption{Comparison of three different methods of estimating the mean energy $E_{\mathrm{n}}$ and the FWHM of the monoenergetic neutron beams (Gaussian fits of the neutron energy spectra with the MINUIT code were used to determine these parameters). The resolution ($Res=FWHM/E_{\mathrm{n}}$) is also shown. The superscripts \textit{exp} and \textit{sim} denote the unfoldings done with the experimental and corrected simulated response matrices, respectively. The label TOF denotes the experimental evaluation done with time of flight. In the cases were TOF results were not available, the results are from the simulation code TARGET and this is indicated with an asterisk.}
\label{tab:res}
\end{table}

The neutron energy spectrum of Fig. \ref{fig:En14MeV} shows the main energy peak at 14 MeV and neutrons from the break-up process at lower energies. This illustrates clearly the good capabilities of the unfolding technique to reveal the whole energy distribution of the field.

Fig \ref{fig:ph14MeV} shows the measured PHS together with the predicted PHS that results from folding the neutron energy spectrum shown in Fig. \ref{fig:En14MeV} with the response functions. There is very good agreement between both PHS.

Finally, a comparison was made between the neutron flux determined by the CNS and by an independent method (based on the PTB accelerator's reference monitor) for a monoenergetic beam with neutron energy $E_{n} = 14$ MeV. The result derived from the PTB monitor is $\varPhi = 168600 ~\mathrm{cm}^{-2}$, while the CNS result is $\varPhi = 166200 ~\mathrm{cm}^{-2}$. The agreement is very good, with a difference of less than $1.5\%$.

\subsection{Comparison between neutron fields}\label{compfields}

\begin{figure}[ht]
\centering
\includegraphics[width=4.5in]{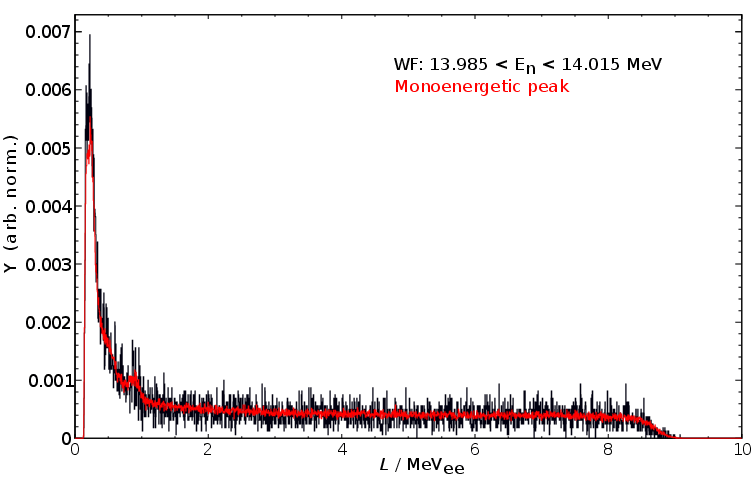}
\caption{Comparison of pulse height spectra from the 14 MeV monoenergetic field (in red) and white field (in black). The energy selection in the white field range from 13.985 MeV to 14.015 MeV.}
\label{fig:wfvsmono}
\end{figure}

Fig. \ref{fig:wfvsmono} shows a comparison between two PHS. The first PHS, in red, is a measurement of a monoenergetic beam with neutron energy $E_{n} = 14$ MeV. The second PHS, in black, was constructed from the measurement of the white field using TOF information to select only those neutrons with energies 13.985 MeV $< E_{\mathrm{n}} <$ 14.015 MeV. The low statistics of the energy slice of the white field explains the large statistical variation observed for the PHS plotted in black.

The excellent agreement between the two PHS provides a check of the whole procedure used for the analysis, in particular of the TOF evaluation that has been implemented. This comparison is important because the experimental response matrix is put together from PHS that are constructed using the energy information obtained from the TOF procedure. Also, different energy slices of the white field measurements were used to evaluate the pulse height resolution.

\section{Conclusions}

We have carried out the characterization of a CNS based on a BC501A fast-neutron detector with a digital acquisition system. To achieve this goal, we developed a new analysis software written in C++ that uses the ROOT framework. The analysis allowed us to determine two response matrices for the system, an experimentally determined response matrix and a response matrix based on simulations adjusted to fit data collected during a measurement campaign carried out at the PTB accelerator facility. We have evaluated both the pulse height resolution and the energy resolution that is achievable after unfolding the measurements with the MAXED code, and compared to the energy resolution determined by TOF methods.

The results obtained for the pulse height resolution are in agreement with values that were obtained previously with similar systems that have larger cell sizes \cite{Klein2002}.

The energy resolution, as determined by the unfoldings of the measurements of monoenergetic beams, shows that a resolution that is comparable to the TOF resolution has been achieved. In particular, the resolution achieved at 2.5 and 14 MeV (see lines 2 and 6 of table \ref{tab:res}) satisfy the requirements listed in section \ref{CNSP}. These results demonstrate the good quality of the response functions and the capabilities of CNS. The energy resolution determined from the unfoldings done with the experimental response matrix are comparable to the energy resolution obtained with TOF, but not as good as that of the unfoldings done with the corrected simulated response matrix.

The comparison between the unfoldings of the white field using the experimental and the corrected simulated response matrices (see Fig. \ref{fig:UnfWF}) show good agreement. There is structure at about 10 MeV in the unfolding done with the corrected simulated response matrix which is not in the unfolding done with the experimental response matrix or in the result derived using TOF. This seems to be connected to the cross section libraries used for the simulation, and it is still under investigation at this time.

\section*{Acknowledgment}
The authors thank A. Kasper, K. Tittelmeier, A. L\"{u}cke, L. Giacomelli and F. Belli for their participation during the experimental campaign at PTB. This work was supported by contract FU06-CT-2006-00076 (EFDA-JET-CT-2005-018).


\begin{thebibliography}{99}

\bibitem{hutch2005}
I.H. Hutchinson, \emph{Principles of plasma diagnostics}, 
    Cambridge University Press, Cambridge 2005.

\bibitem{Klein2002}
H. Klein and S. Neumann,
\emph{Neutron and photon spectrometry with liquid scintillation detectors in mixed fields}, in proceedings of
\emph{Int. Workshop on Neutron Field Spectrometry in Science, Technology and Radiation Protection}, June 4-8, 2008 Pisa, Italy,
\href{http://dx.doi.org/10.1016/S0168-9002(01)01410-3}
{\emph{Nucl. Inst. \& Meth. in Phys. Res. A} {\bf 476} (2002) 132}.

\bibitem{Zimbal2004}
A. Zimbal et al.,
\emph{Compact NE213 neutron spectrometer with high energy resolution for fusion applications},
\href{http://dx.doi.org/10.1063/1.1787935}
{\emph{Rev.\ Sci.\ Instrum.} {\bf 75} (2004) 3553}.

\bibitem{Reginatto2008}
M. Reginatto and A. Zimbal,
\emph{Bayesian and maximum entropy methods for fusion diagnostic measurements with compact neutron spectrometers},
\href{http://dx.doi.org/10.1063/1.2841695}
{\emph{Rev.\ Sci.\ Instrum.} {\bf 79} (2008) 023505}.

\bibitem{Tardini2012}
G. Tardini et al., \emph{First neutron spectrometry measurements in the ASDEX Upgrade tokamak}, in proceedings of this conference.

\bibitem{Nolte2004}
R. Nolte et al.,
\emph{Quasi-monoenergetic neutron reference fields in the energy range from thermal to 200 MeV},
\href{http://dx.doi.org/10.1093/rpd/nch195}
{\emph{Radiat. Prot. Dosim.} {\bf 110} (2004) 97}.

\bibitem{Moisan2012}
F. Gagnon-Moisan, A. Zimbal and M. Reginatto, \emph{New developments for the determination of the response function for a BC501A compact neutron spectrometer for fusion diagnostics}, in proceedings of
\emph{Advancements in Nuclear Instrumentation, Measurement Methods and their Applications}, June, 6--9, 2011 ICC, Ghent, Belgium
\emph{submitted}.

\bibitem{root}
R. Brun and F. Rademakers,
\emph{ROOT-an object oriented data analysis framework}, in proceedings of
\emph{AIHENP'96 Workshop}, September, 1996 Lausanne, France,
\emph{Nucl. Inst. \& Meth. in Phys. Res. A} {\bf 389} (1997) 81,
See also http://root.cern.ch/.

\bibitem{Marocco2009}
D. Marocco et al.,
\emph{High count rate neutron spectrometry with liquid scintillation detectors},
\href{http://dx.doi.org/10.1109/TNS.2009.2020164}
{\emph{IEEE Trans. Nucl. Sci.} {\bf 56} (2009) 1168}.

\bibitem{Esposito2004}
B. Esposito et al.,
\emph{Neutron measurements on Joint European Torus using an NE213 scintillator with digital pulse shape discrimination},
\href{http://dx.doi.org/10.1063/1.1785278}
{\emph{Rev.\ Sci.\ Instrum.} {\bf 75} (2004) 3550}.

\bibitem{Roush1964}
M.L. Roush, M.A. Wilson and W.F. Hornyak,
\emph{Pulse shape discrimination},
\href{http://dx.doi.org/10.1016/0029-554X(64)90333-7}
{\emph{Nucl. Inst. \& Meth.} {\bf 31} (1964)112}.

\bibitem{Knoll2010}
G.F. Knoll, \emph{Radiation detection and measurements}, John Wiley
    and Sons, Inc., New York 2000.

\bibitem{Brooks2002}
F.D Brooks and H. Klein,
\emph{Neutron spectrometry-historical review and present status}, in proceedings of
\emph{Int. Workshop on Neutron Field Spectrometry in Science, Technolog y and Radiation Protection}, June 4-8, 2008 Pisa, Italy,
\href{http://dx.doi.org/10.1016/S0168-9002(01)01378-X}
{\emph{Nucl. Inst. \& Meth. in Phys. Res. A} {\bf 476} (2002) 1}.

\bibitem{NRESP}
G. Dietze and H. Klein, \emph{Montecarlo Codes for the calculation of neutron response functions and detection efficiencies for NE213 scintillation detectors}, Physikalisch-Technische Bundesanstalt Report PTB-ND-22, Braunschweig, Germany 1982.

\bibitem{Zimbal2006}
A. Zimbal, H. Klein, M. Reginatto, H. Schuhmacher, L. Bertalot and A. Murari, \emph{High resolution neutron spectrometry with liquid scintillation detectors for fusion applications}, in proceedings of
\emph{International workshop on fast neutron detectors and
applications}, April, 3--6, 2006 University of Cape Town, South Africa, 
\pos{PoS(FNDA2006)035}.

\bibitem{Schmidt1998}
D. Schmidt and H. Klein, \emph{Precise time-of-flight spectrometry of fast neutrons principles, methods and results}, Physikalisch-Technische Bundesanstalt Report PTB-N-35, Braunschweig, Germany 1998.

\bibitem{Bardelli2007}
L. Bardelli, G. Poggi, M. Bini, G. Pasquali, NUCL-EX and FAZIA Collaborations,
\emph{An efficient method for timing synchronization between many digital sampling channels},
{\emph{Nucl. Inst. \& Meth. in Phys. Res. A} {\bf 572} (2007) 882}.

\bibitem{Novotny1997}
T. Novotn\'{y}, \emph{Photon spectrometry in mixed neutron-photon fields using NE 213 liquid scintillation detectors}, Physikalisch-Technische Bundesanstalt Report PTB-N-28, Braunschweig, Germany 1997.

\bibitem{Linden1995}
W. von der Linden,
\emph{Maximum-entropy data analysis },
\href{http://dx.doi.org/10.1007/BF01538241}
{\emph{Appl. Phys. A} {\bf 60} (1995) 155}.

\bibitem{Sivia2006}
D. Sivia and J. Skilling, \emph{Data analysis: a Bayesian tutorial}, Oxford University Press, New York 2006.

\bibitem{fletcher}
R. Fletcher, \emph{Practical methods of optimization}, John Wiley
    and Sons, Inc., New York 2000.

\bibitem{GSL}
M. Galassi et al, \emph{GNU Scientific Library reference manual-third edition}, Network Theory Ltd., Bristol 2009, See also http://www.gnu.org/software/gsl/.

\bibitem{TARGET}
D. Schlegel, \emph{Target user's manual}, Physikalisch-Technische Bundesanstalt Report PTB-642-05-2, Braunschweig, Germany 2005.

\end{thebibliography}
\end{document}